\documentclass[twocolumn,aps,pra,superscriptaddress]{revtex4-1}
\usepackage{graphicx}
\usepackage{amsmath}
\usepackage{amssymb}
\usepackage{braket}
\usepackage[colorlinks=true,linkcolor=blue,citecolor=blue,urlcolor=blue]{hyperref}

\begin{document}

\title{Asymmetric Conductivity of the Kondo Effect in Cold Atomic Systems}

\author{Yanting Cheng}
\affiliation{Institute for Advanced Study, Tsinghua University, Beijing, 100084,
China}

\author{Xin Chen}
\affiliation{Institute for Advanced Study, Tsinghua University, Beijing, 100084,
China}

\author{Ren Zhang}
\email{renzhang@xjtu.edu.cn}
\affiliation{School of Science, Xi'an Jiaotong University, Xi'an, 710049, China}

\date{\today}
\begin{abstract}
Motivated by recent theoretical and experimental advances in quantum simulations using alkaline earth(AE) atoms, we put forward a proposal to detect the Kondo physics in a cold atomic system. It has been demonstrated that the intrinsic spin-exchange interaction in AE atoms can be significantly enhanced near a confinement-induced resonance(CIR), which facilitates the simulation of Kondo physics. Since the Kondo effect appears only for antiferromagnetic coupling, we find that  the conductivity of such system exhibits an asymmetry across a resonance of spin-exchange interaction. The asymmetric conductivity can serve as the smoking gun evidence for Kondo physics in the cold atom context. When an extra magnetic field ramps up, the spin-exchange process near Fermi surface is suppressed by Zeeman energy and the conductivity becomes more and more symmetric.  Our results can be verified in the current experimental setup.
\end{abstract}
\maketitle

\section{ Introduction}
In past decades, cold atom physics have achieved a plenty of remarkable triumphs on quantum simulation, including fermionic pairing, Bose and Fermi Hubbard model, topological phase and topological transition, et. al.\cite{qs1,qs2,qs3}. Meanwhile, thanks to the high controllability of cold atom, lots of new universal physics have been unveiled apart from that established in the condensed matter counterpart, such as BCS-BEC crossover and quench dynamics and so on\cite{exp1,exp2,exp3,exp4,exp5,exp6,exp7}. As a result, cold atom system has become an ideal platform to discover new physics even in the well-established model, such as the Kondo model\cite{stoof2004,Gorshkov2010,Foss2010,Bauer2013,nishida2013,Isaev2015,kuzmenko2015,nishida2016,ZR2016,Kuzemenko2018,1dcir,Yao2018}.

Kondo effect refers to the anomalous increase of resistivity below a particular temperature (Kondo temperature) in a metal with magnetic impurities. In this system, a spatially localized magnetic impurity is embedded in a cloud of electronic gas, and the spins of impurity and itinerant electrons can be exchanged via collision.
According to the poor man scaling scenario\cite{poor}, in the low-temperature regime, the lower is the temperature, the stronger is the effective antiferromagnetic spin interaction, which favors the Kondo effect. However, for the ferromagnetic spin interaction, the situation is significantly different and the Kondo effect does not exist. A pioneering work by Kondo\cite{ori} discovered that the third-order perturbation is crucial for the understanding of Kondo effect. Specifically, the third-order perturbation shows that the resistivity is logarithmically dependent on the temperature below Kondo temperature, which has become the hallmark of the Kondo effect.

In order to simulate the Kondo model in cold atom system, several requirements have to be satisfied. First of all, two species atoms should have very different ac polarization. Hence, one of them can be spatially localized by a deep optical lattice playing the role of impurity while the other one retains its mobility. Secondly, there has to be a spin-exchange interaction between the impurity and itinerant atoms. At last, if the spin-exchange interaction is weak, the Kondo temperature will be orders of magnitude lower than the Fermi temperature, which is hardly achievable in current cold atom experiment. For this reason, the spin-exchange interaction should be significantly strong so as to enhance the Kondo temperature.

Alkaline-earth(AE) atom features the coexistence of a stable ground state $^{1}S_{0}$ (labeled as $|g\rangle$) and a meta-stable clock state $^{3}P_{0}$ (labeled as $|e\rangle$) which can be defined as the orbit degree of freedom. The ac polarization for these two orbital states is generally different except for the laser with a magic wavelength\cite{mag1,mag2}. In the presence of a particular laser, the $|e\rangle$-atom can be spatially localized while the $|g\rangle$-atoms are freely moving. Thereupon, the $|e\rangle$-atom plays the role of impurity and the $|g\rangle$-atoms play the role of the itinerant majority. Furthermore, the nuclear spin of the fermionic AE atom is nonzero and it has been demonstrated that there is an intrinsic spin-exchange interaction between the ground state and the clock state\cite{spinexchange1,spinexchange2}. To one's delight, both theoretical and experimental works have verified that the spin-exchange interaction can be resonantly enhanced by a confinement-induced resonance(CIR)\cite{1dcir,1dcirex,Zhang2018}. On these accounts, cold AE atoms system has become the most promising platform to explore the Kondo physics. 

Up to now, the problem demanding an urgent solution is how to detect the Kondo effect in cold atoms if it were realized. The prevailing wisdom for detecting the Kondo effect in magnetic metal is to measure the resistivity of the itinerant electrons with various temperature\cite{kondo}. However, it is difficult to perform such a measurement in cold atom experiment. In this manuscript, we addressed this problem by studying the transport properties of one dimensional Kondo model with a tunable coupling strength. Our results show that the conductivity of itinerant atoms exhibits an obvious asymmetry with respect to the location of a resonance of spin exchange interaction, which can serve as a smoking gun for Kondo effect in this system. When an extra magnetic field is ramped up, the conductivity becomes more symmetric, since the possibility of  spin exchange process near Fermi surface is suppressed by the Zeeman energy.

\begin{figure}
\centering
\includegraphics[width=0.45\textwidth]{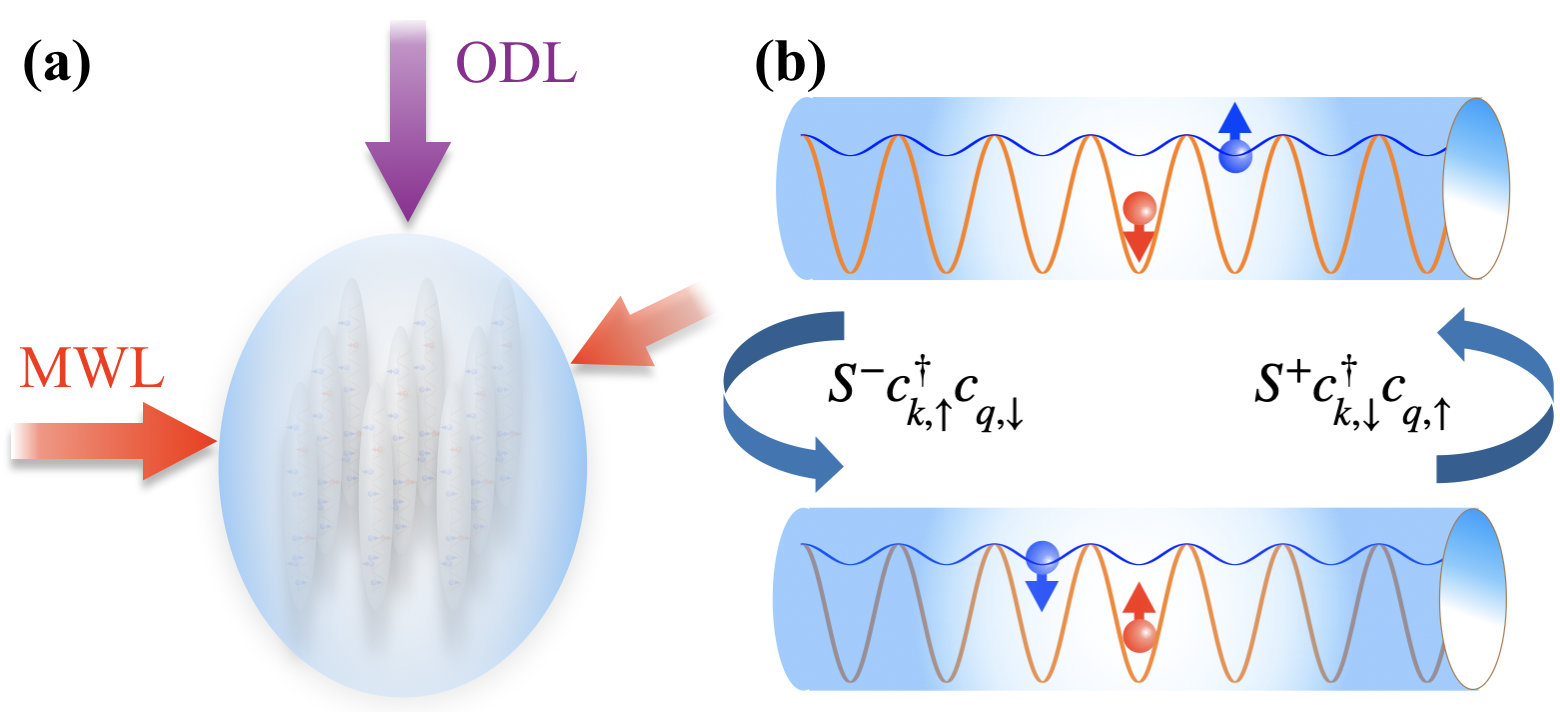}
\caption{Illustration of the spin-exchange interaction between impurities and itinerant atoms. For instance, in $^{173}$Yb Fermi gas, blue and red ball denote the ground state $^{1}S_{0}$ and clock state $^{3}P_{0}$ atom, respectively. The arrows associated with atoms are two of the $N$ components of nuclear spin. (a): Both $^{1}S_{0}$ and $^{3}P_{0}$ state atoms are simultaneously trapped by magic wavelength laser(MWL) in the transverse direction.  Shined by an orbit-dependent laser(ODL) in the axial direction, $^{3}P_{0}$ state atoms experience a deep lattice, and hence are spatially localized, while $^{1}S_{0}$ state atoms feel a much shallower lattice and still keep its mobility. (b): The nuclear spin of these two states can be exchanged via collision.}
\label{schematic}
\end{figure}
\section{Conductivity of one-dimensional Kondo model.}
Inspired by recent experiments\cite{1dcirex}, we start with the one dimensional Kondo system schematized in Fig.\ref{schematic}, which can be described  by a 1D Hamiltonian $\hat{H}_{\rm 1D}=\hat{H}_{0}+\hat{H}_{\rm{int}}$ with
\begin{align}
\hat{H}_{0}=&-\frac{1}{N_{s}}\sum_{\vec{k},\sigma=\uparrow,\downarrow}\epsilon_{k} \hat{c}^\dagger_{\vec{k},\sigma}\hat{c}_{\vec{k},\sigma},\\
\label{effect-Kondo}
\hat{H}_{\rm{int}}=&\frac{J_1}{N_s}\sum_{\vec{k},\vec{k}'}\hat{\textbf{S}}\cdot\hat{\textbf{s}}_{\vec{k}\vec{k}'}+\frac{J_2}{N_s}\sum_{\vec{k},\vec{k}',\sigma=\uparrow,\downarrow}n_{\rm{imp}}\hat{n}_{\vec{k}\vec{k}'\sigma}.
\end{align}
where the dispersion of the itinerant atoms reads $\epsilon_{k}=-2t\cos(ka)$ with $t$ denoting the hopping strength of itinerant atoms, $a$ is the lattice constant, $J_1$ and $J_2$ are coupling parameters which can be tuned in current experiment, $N_s$ denotes the site number, $\hat{\textbf{S}}$ and $\hat{\textbf{s}}_{\vec{k}\vec{k}'}$ denote the spin operator for impurities and itinerant atoms respectively ($\hat{s}_{\vec{k}\vec{k}'}^+=\hat{c}_{\vec{k}\uparrow}^\dagger\hat{c}_{\vec{k}'\downarrow}$ and $\hat{s}_{\vec{k}\vec{k}'}^-=\hat{c}_{\vec{k}\downarrow}^\dagger\hat{c}_{\vec{k}'\uparrow}$), $n_{\rm_{imp}}$ is the impurity concentration, and $\hat{n}_{\vec{k}\vec{k}'\sigma}=\hat{c}_{\vec{k}\sigma}^\dagger\hat{c}_{\vec{k}'\sigma}$ is the itinerant atoms density. Here, we would like to point out that different from the original Kondo model, there is a density-density interaction described by the second term in Eq.(\ref{effect-Kondo}).  The expected phase diagram for this system should be more rich.

Utilizing the Kubo formula, one can straightforwardly obtain the direct conductivity of the itinerant atoms. In the dilute impurity limit, it can be expressed as\cite{dia}
\begin{align}
\label{kubo}
\lim_{n_{\rm{imp}}\rightarrow0}\sigma=\frac{2}{m^2}\int\frac{d\vec{p}}{(2\pi)}p^2\tau(p)\left[-\frac{\partial n_F(\epsilon_p)}{\partial\epsilon_p}\right],
\end{align}
where $m$ denotes the atomic mass, $n_{F}(\epsilon_{p})$ is the Fermi distribution, and the relaxation $\tau(p)$ is defined as following
\begin{align}
\label{relaxiationT}
\frac{1}{\tau(p)}=2\pi n_{\rm imp}\int\frac{d\vec{p'}}{2\pi}|T_{\vec{p}\vec{p}'}|^2\delta(\epsilon_{p}-\epsilon_{p'})\left(1-\frac{\vec{p}\cdot{\vec{p}}'}{p^2}\right)
\end{align}
with $|T_{\vec{p}\vec{p}'}|$ denoting the determinant of the scattering $T$-matrix element between the Fermi sea of  the itinerant atoms and the localized impurity atom. The energy conservation is manifested by the presence of $\delta$-function. We would like to emphasis that $\partial n_{F}(\epsilon_{p})/\partial \epsilon_{p}$ factor in Eq.(\ref{kubo}) indicates only these itinerant atoms near the Fermi surface contribute significantly to conductivity.

It is known that for a many-body system, the exact $T$-matrix is generally formidable. However, if the qualitative behavior of the conductivity is concerned, one can use the prevailing perturbation theory to calculate the $T$-matrix elements in Eq.(\ref{relaxiationT}).  
In our calculation, we assume that the itinerant atoms are half-filling and only consider the scattering between the impurity atom and the itinerant atoms near the Fermi surface.
A straightforward calculation gives the the first-order of $T$-matrix elements as follows
\begin{subequations}
\label{T1}
\begin{align}
&\bra{\vec{p},\uparrow}T\ket{\vec{p'},\uparrow}_{(1)}=\frac{J_1}{N_s}S^z+\frac{J_2}{N_s},\\
&\bra{\vec{p},\downarrow}T\ket{\vec{p'},\downarrow}_{(1)}=-\frac{J_1}{N_s}S^z+\frac{J_2}{N_s},\\
&\bra{\vec{p},\uparrow}T\ket{\vec{p'},\downarrow}_{(1)}=\frac{J_1}{N_s}S^-,\\
&\bra{\vec{p},\downarrow}T\ket{\vec{p'},\uparrow}_{(1)}=\frac{J_1}{N_s}S^+.
\end{align}
\end{subequations}
One can see that the leading order of $|T_{\vec{p}\vec{p}'}|^2\propto J_{1}^{2},J_{2}^{2}$, hence is independent on the sign of $J_{1}$. It makes sense because one has to resort to the third order perturbation to manifest the Kondo effect. If the density-density interaction is turned off, i.e., $J_{2}=0$, Eq.(\ref{T1}) will recover the results given by the original Kondo model.

\begin{figure}
\centering
\includegraphics[width=0.3\textwidth]{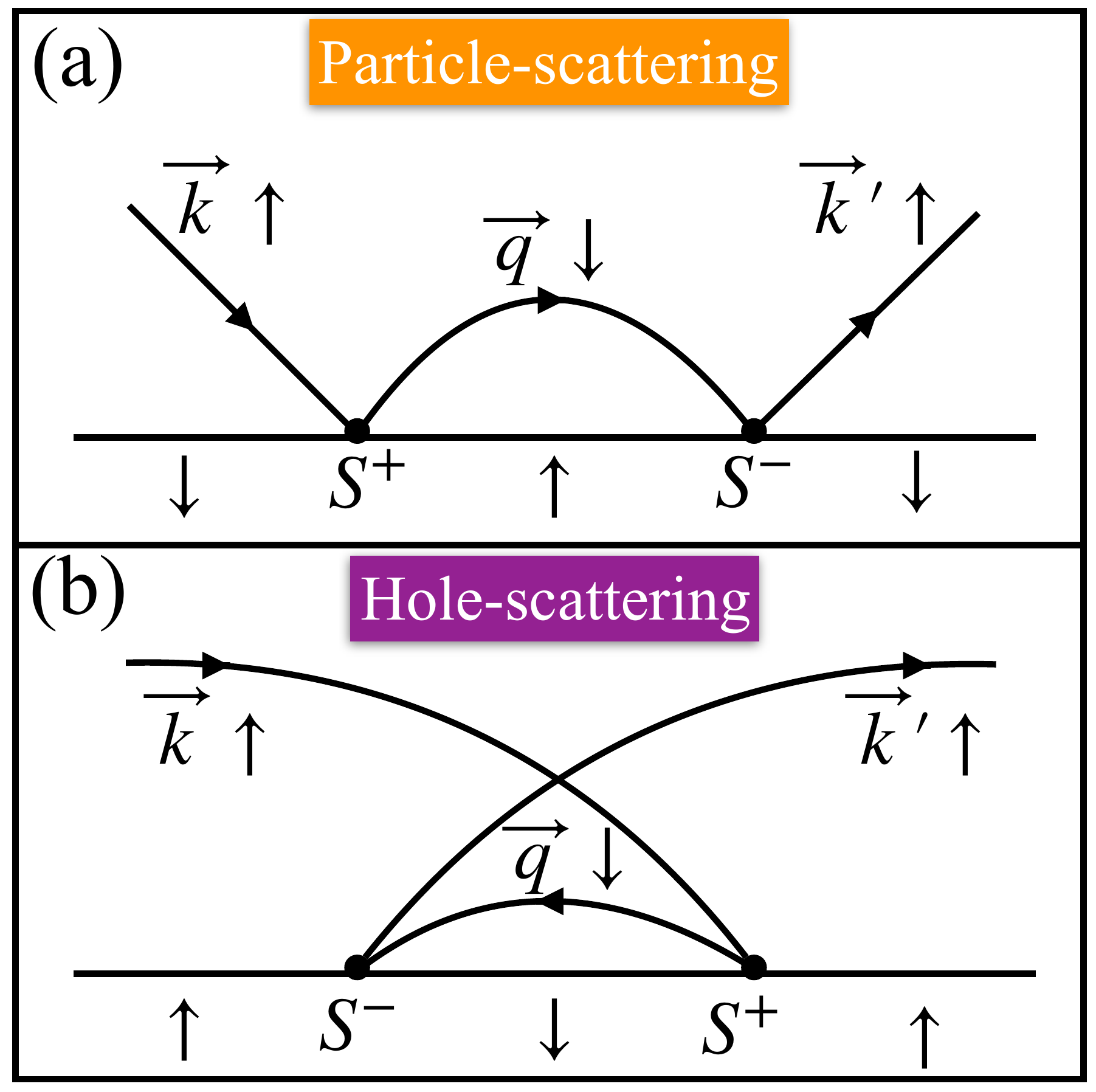}
\caption{Two typical second-order contributions to the $T$-matrix. $S^+$ and $S^-$ are spin operators for impurities. (a): Contribution by extracting a particle from the Fermi sea. A spin up incident atom with momentum $\vec{k}$ interacts with an impurity by exchanging spin, thus it becomes spin down with momentum $\vec{q}$. And then it interacts with the same impurity by exchanging spin again, at last it keeps spin up and emit with momentum $\vec{k}'$. (b): Contribution by occupying a hole in the Fermi sea. A spin up atom with momentum $\vec{k}'$ and a spin down with momentum $\vec{q}$ hole are created, and this atom emit. Then a spin up incident atom with momentum $\vec{k}$ annihilate with this hole.}
\label{2ndT}
\end{figure}

Compared to the first-order $T$-matrix, the derivation of second-order $T$-matrix is much more involved. In FIG.\ref{2ndT}, we show two typical scattering processes which contribute to the second-order $T$-matrix elements. It should be noticed that when the magnetic field is present,  the degeneracy of the itinerant atoms with different spin will be lifted by the Zeeman energy. Here we adopt  $\epsilon_{k\uparrow}=-2t\cos(ka)+h$ and $\epsilon_{k\downarrow}=-2t\cos(ka)-h$ to distinguish dispersions of itinerant atoms with $h$ being the Zeeman energy. 
Since only scattering taking place near the Fermi surface contributes to the $T$-matrix, it is reasonable to perform the approximation of $k=k'\approx k_{F}$ with $k$ and $k'$ denote the incident and outing momentum of itinerant atom and $k_{F}$ is the Fermi momentum. By setting the zero energy as the Fermi energy, one can  straightforwardly  obtain the contributions of processes illustrated in FIG.\ref{2ndT}(a),(b), which can be explicitly expressed as
\begin{subequations}
\begin{align}
&\frac{J_{1}^{2}}{N_{s}^{2}}\sum_{\vec{q}}S^{+}S^{-}\frac{n_{F}(\epsilon_{\vec{q},\downarrow})}{\epsilon_{p}+i\delta-\epsilon_{\vec{q},\downarrow}},\\
&\frac{J_{1}^{2}}{N_{s}^{2}}\sum_{\vec{q}}S^{-}S^{+}\frac{1-n_{F}(\epsilon_{\vec{q},\downarrow})}{\epsilon_{p}+i\delta-\epsilon_{\vec{q},\downarrow}},
\end{align}
\end{subequations}
respectively.
Summing up all of the possible contributions and using the relations that $S^{-}S^{+}=1/2-S^{z}, S^{+}S^{-}=1/2+S^{z},S^{-}S^{z}=S^{-}/2,S^{z}S^{-}=-S^{-}/2,S^{+}S^{z}=S^{+}/2,S^{z}S^{+}=S^{+}/2$, one finally obtains the second-order $T$-matrix elements
\begin{subequations}
\label{T2}
\begin{align}
\bra{\vec{k},\uparrow}T\ket{\vec{k}',\uparrow}_{(2)}=&S^z\left[g_{\downarrow}(\epsilon_{p})+S^{z}\delta n(\epsilon_{p})\right],\\
\bra{\vec{k},\downarrow}T\ket{\vec{k}',\downarrow}_{(2)}=&S^z\left[g_{\uparrow}(\epsilon_{p})+S^{z}\delta n(\epsilon_{p})\right],\\
\bra{\vec{k},\uparrow}T\ket{\vec{k}',\downarrow}_{(2)}=&S^-\left[\frac{g_{\uparrow}(\epsilon_{p})+g_{\downarrow}(\epsilon_{p})}{2}\right],\\
\bra{\vec{k},\downarrow}T\ket{\vec{k}',\uparrow}_{(2)}=&S^+\left[\frac{g_{\uparrow}(\epsilon_{p})+g_{\downarrow}(\epsilon_{p})}{2}\right].
\end{align}
\end{subequations}
where $I$ denotes the $2\times2$ identity matrix. $g_{\uparrow(\downarrow)}(\epsilon_{p})$ and $\delta n(\epsilon_{p})$ are defined as
\begin{align}
g_{\uparrow(\downarrow)}(\epsilon_{p})&\equiv\frac{J^{2}_1}{N^{2}_s}\sum_{\vec{q}}\frac{2n_{F}(\epsilon_{\vec{q},\uparrow(\downarrow)})}{\epsilon_{p}+i\delta-\epsilon_{\vec{q},\uparrow(\downarrow)}},\\
\delta n(\epsilon_{p})&\equiv\frac{J^{2}_1}{N^{2}_s}\sum_{\vec{q}}\frac{n_{F}(\epsilon_{\vec{q},\uparrow})-n_{F}(\epsilon_{\vec{q},\downarrow})}{\epsilon_{p}+i\delta},
\end{align}
respectively.
Here, we would like to point out that since only the states near Fermi surface contribute to the conductivity, and the itinerant atoms are half-filling, the contribution of all the terms containing $\sum_{\vec{q}}(\varepsilon+i\delta-\epsilon_{\vec{q},\uparrow(\downarrow)})^{-1}$ has been ignored in our derivation. Collecting Eq.(\ref{T2}) and Eq.({\ref{T1}}) together and substituting them into Eq.(\ref{relaxiationT}), one can obtain the conductivity of the itinerant atom which can be the measured quantity under current experiment condition.

\section{Asymmetry of the conductivity.}
Due to spin-exchange interaction, the resistivity of magnetic metal is $\log T$-dependent in the low-temperature regime. While, the phonon excitation leads to a $T^{5}$-dependence in the the high-temperature regime, As a result, the resistivity takes its minimum at the Kondo temperature\cite{kondo}.  The conventional wisdom for detecting the Kondo effect in condensed matter counterpart is to measure the temperature dependency of the resistivity. However, it is difficult to perform such a measurement in cold atom system. First of all, since there is no phonon excitation in the optical lattice, the $T^{5}$-dependency of conductivity is absent in cold atom system. On the other hand, the available temperature window of Kondo effect in cold atoms is not clear yet. In the present manuscript, we propose that the Kondo effect can be manifested by the asymmetric conductivity of the itinerant atoms across a resonance of the spin-exchange interaction.

Kondo effect only takes place in the antiferromagnetic regime. In this regime, as the temperature decreases the effective spin-exchange interaction becomes stronger and stronger. The spin of the impurity will be screened by itinerant atomic cloud, which leads to an enhancement of the itinerant atom scattering. Hence, the conductivity can be significantly suppressed. This picture, interestingly, will be distorted in the ferromagnetic regime where the Kondo effect does not exist. Hence there is no suppression for the conductivity anymore.  In cold atoms, one can take the advantage of scattering resonance technique to adiabatically drive the system from ferromagnetic regime to antiferromagnetic regime.  The expected conductivity should be asymmetric with respect to the location of the resonance as long as the temperature is lower than the Kondo temperature.

\begin{figure}
\includegraphics[width=0.42\textwidth]{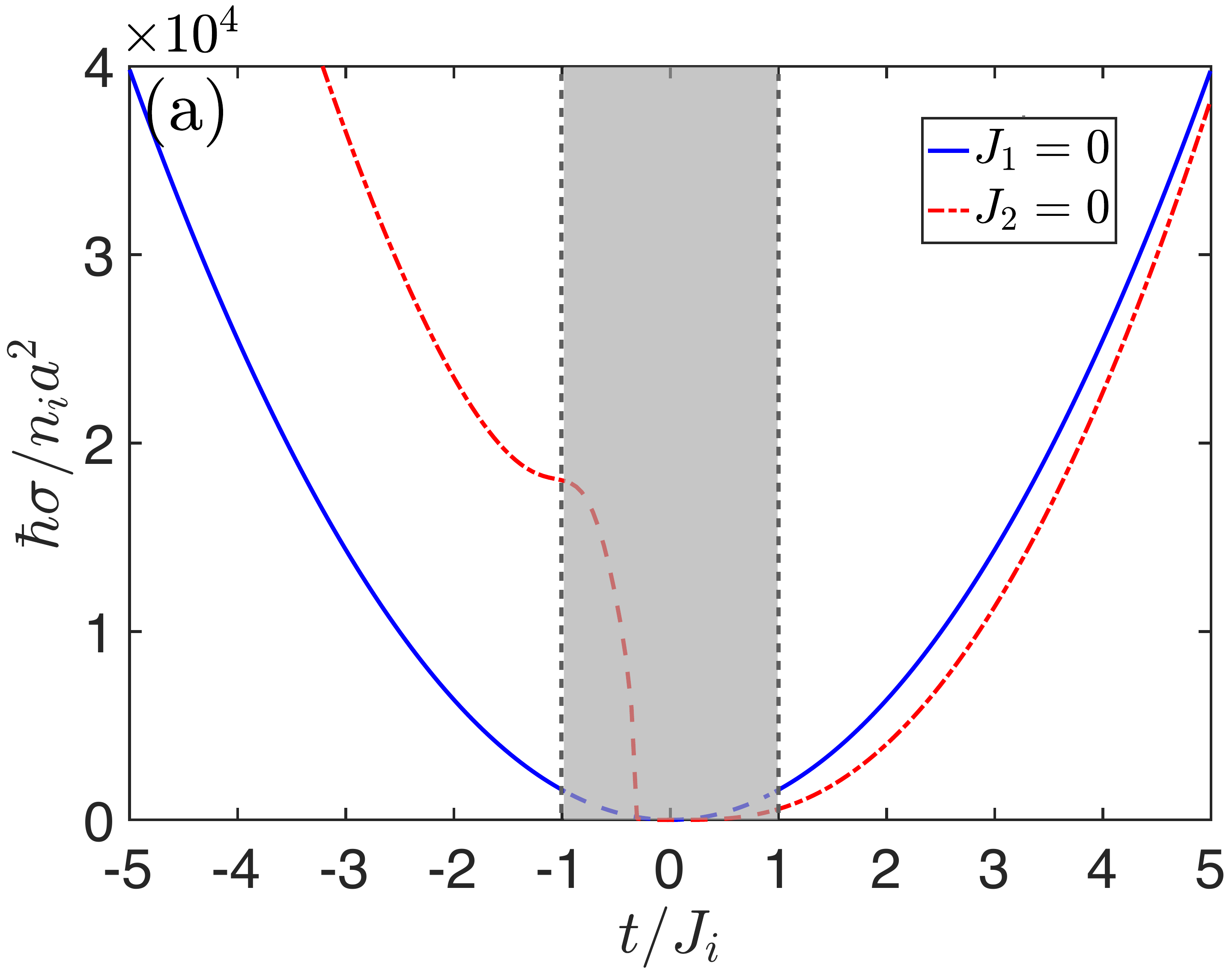}
\includegraphics[width=0.47\textwidth]{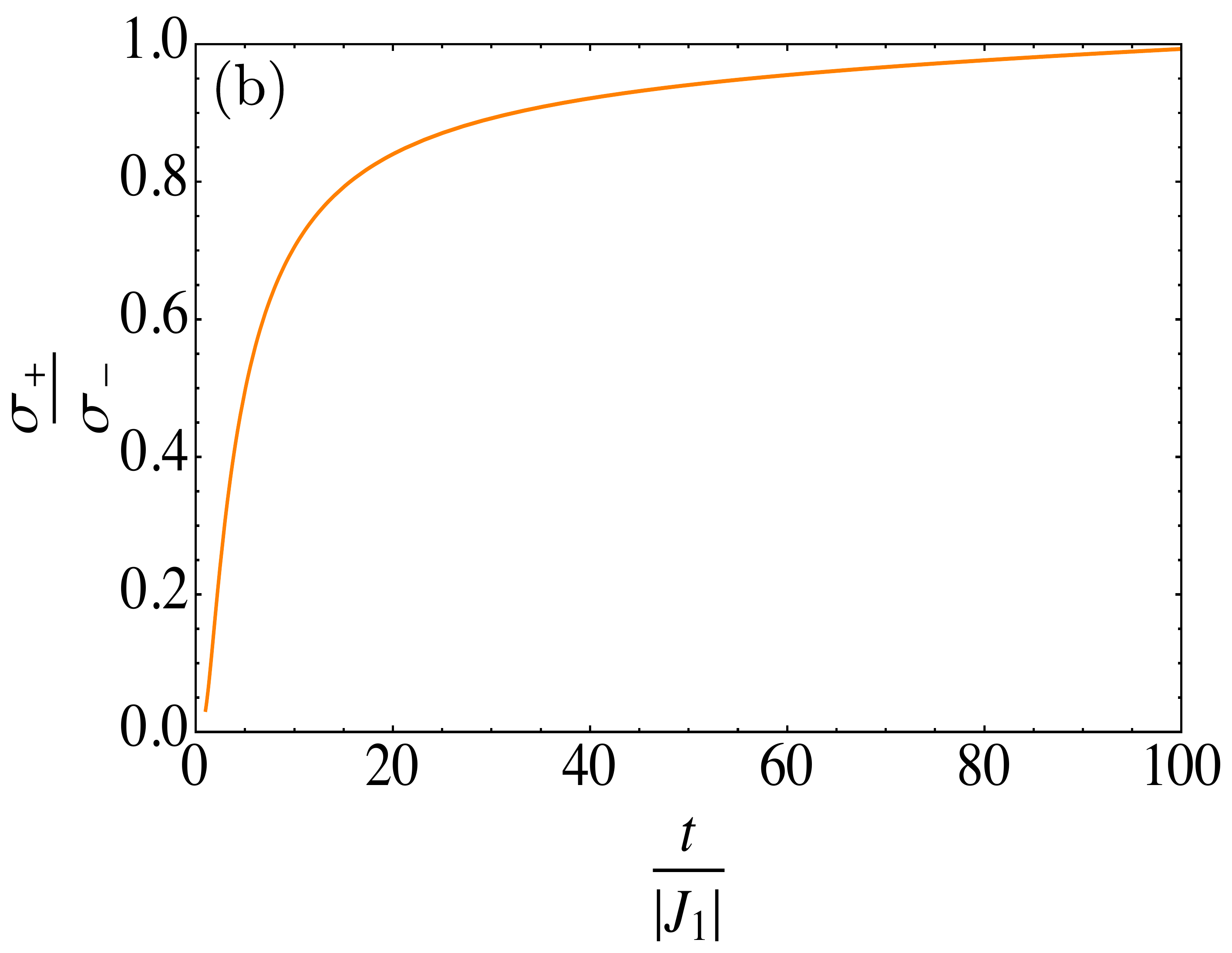}
\caption{(a): Conductivity of itinerant atoms across a resonance of the coupling strength with $n_{\rm imp}=5\times10^{-4}n$, $k_BT=0.2t$ and $h=0$. When the spin-exchange interaction is turned off ($J_1=0$), the conductivity across non-spin-exchange resonance is symmetric as shown by the blue solid curve. When the non-spin-exchange interaction is turned off ($J_{2}=0$), the conductivity across the spin-exchange resonance exhibits an obvious asymmetry. The suppression of conductivity in the antiferromagnetic regime is due to the Kondo effect. The perturbation theory is not applicable to the gray region. (b): The asymmetry $\sigma^{+}/\sigma^{-}$ as function of the reciprocal coupling strength $t/|J_1|$. The asymmetry is quite prominent ($\sigma^{+}/\sigma^{-}\ll1$) near the resonance and eventually disappear ($\sigma^{+}/\sigma^{-}\to1$) when the spin-exchange interaction is too weak.} 
\label{asymmetry}
\end{figure}

In FIG.\ref{asymmetry}, we present the conductivity of the itinerant atoms across a resonance of spin-exchange interaction. As a benchmark, we firstly turn off the spin-exchange interaction($J_{1}=0$) and only keep the non-spin-exchange interaction ($J_{2}\neq0$). In such case, there is no Kondo effect and the conductivity is symmetric with respect to the location of the non-spin-exchange interaction resonance as shown by the blue solid curve in FIG.\ref{asymmetry}(a).  However, if the spin-exchange interaction is turned on($J_{1}\neq0$), the Kondo effect will appear in the antiferromagnetic regime. Our results show that in such case the conductivity exhibits an explicit asymmetry. Compared to that in the ferromagnetic regime($J_{1}<0$), conductivity in the antiferromagnetic regime($J_{1}>0$) is strongly suppressed. The underlying reason for such phenomenon is that the term proportional to $J_{1}^{3}$ plays an essential role in the Kondo effect. In other words, the sign of the spin-exchange interaction can lead to a significant difference of conductivity for the same interaction strength $|J_1|$. We would like to point out that the perturbation theory is valid in the regime of $|t/J_{1,2}|>1$, so our results are reliable in this regime. Nevertheless, at the resonance point $t/J_{1,2}\to0$, our results are reasonable as well, since the impurity is impenetrable in the Tonks-Girardeau gas limit, which means that the conductivity should decay to zero. 

The asymmetry of the conductivity can be quantified by the ratio of $\sigma^{+}/\sigma^{-}$, where $\sigma^{+}\equiv\sigma(|J_{1}|)$ and $\sigma^{+}\equiv\sigma(-|J_{1}|)$ symbolize the conductivity in the antiferromagnetic and ferromagnetic regime. In FIG.\ref{asymmetry}(b), we present the $\sigma^{+}/\sigma^{-}$ as function of $t/|J_{1}|$. It is clear that near the resonance, the asymmetric is very prominent, $\sigma^{+}/\sigma^{-}\ll1$. While when $|J_{1}|$ becomes smaller and smaller, $\sigma^{+}/\sigma^{-}\to1$, which means the symmetric conductivity restores eventually.

\begin{figure}[h]
\includegraphics[width=0.45\textwidth]{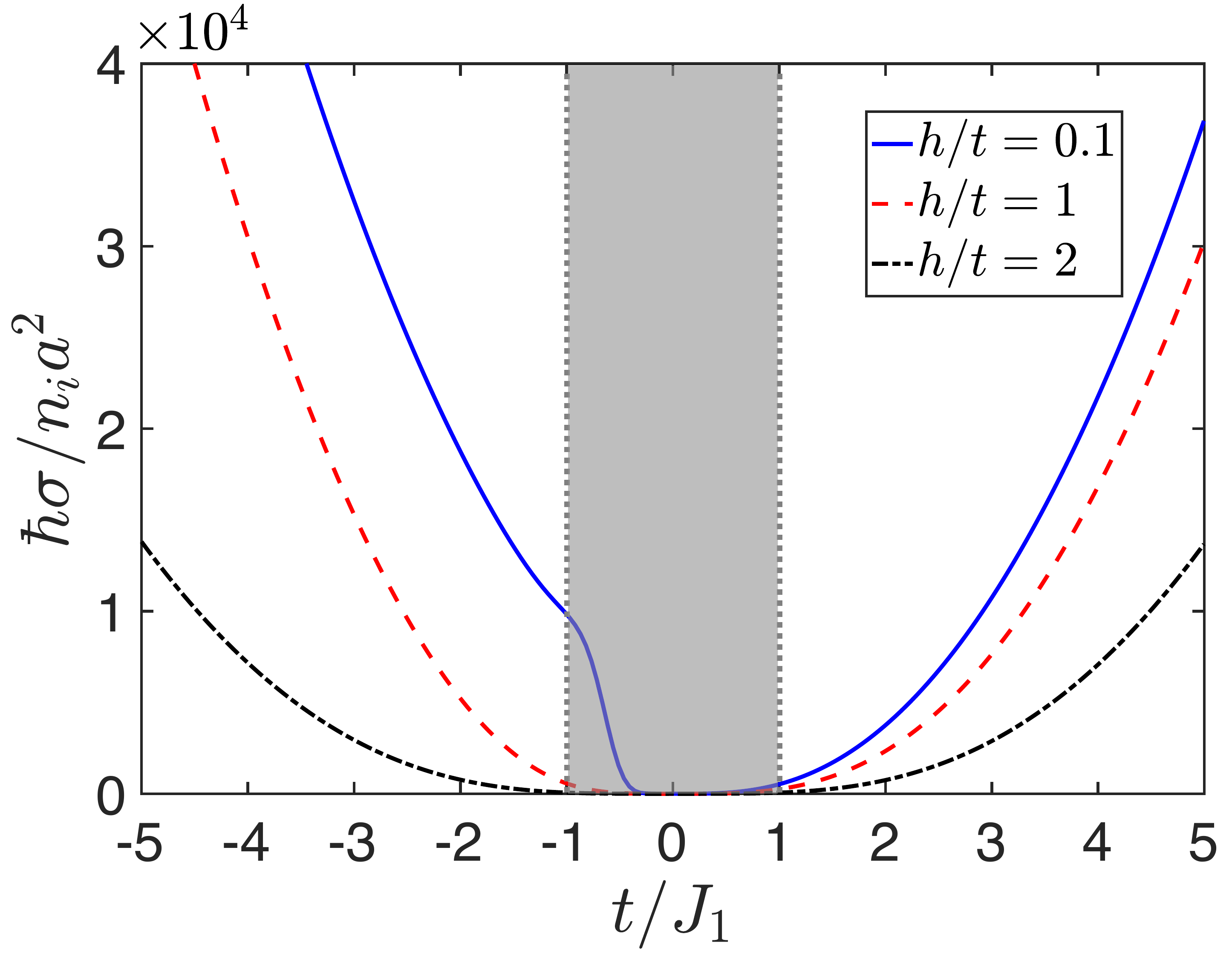}
\caption{The conductivity of itinerant atoms near a resonance with magnetic field. In the presence of magnetic field, the possibility of spin-exchange process is suppressed by the Zeeman energy, which is detrimental to the Kondo effect. Thus, the symmetry of the conductivity restores as the magnetic field ramps up.  In our calculation, we set $J_{2}=0$ for simplicity and the parameters of $n_{\rm imp}=5\times10^{-4}n$, $k_BT=0.2t$ are adopted.}
\label{finite-h}
\end{figure}

We apply an extra magnetic field to demonstrate that the asymmetry of conductivity results from spin exchange interaction. As shown in FIG.\ref{finite-h}, the symmetry of conductivity restores gradually when the Zeeman energy increases. In the presence of a magnetic field,  the itinerant atom is partially polarized. In other words, the density-of-state near Fermi sea for one of the spin components decreases dramatically. As a consequence, the possibility for the spin-exchange is suppressed, which is detrimental to the  Kondo effect. If the magnetic field is so strong that the itinerant atoms are totally polarized, there will be no spin-exchange process and the Kondo effect will disappear. Hence the conductivity becomes symmetric as shown by the black dash-dot curve in FIG.\ref{finite-h}. In our calculation, we set $J_{2}=0$ for simplicity.

\section{Summary and Outlook.}
In summary, by studying transport properties, we find the conductivity of itinerant atoms in one dimension Kondo model exhibits an asymmetric behavior across a resonance of coupling strength, which can serve as a smoking gun for the Kondo effect in this system. Recent works have demonstrated that the spin-exchange interaction can be tuned by a CIR, hence the spin-exchange interaction can be antiferromagnetic or ferromagnetic. In the antiferromagnetic regime, the conductivity of the itinerant atoms can be strongly suppressed by the Kondo effect, which gives rise to the asymmetric conductivity. In the presence of a magnetic field, the possibility for spin-exchange process near Fermi surface is suppressed by the Zeeman energy, and the symmetric conductivity of itinerant atom restores gradually when the magnetic field becomes stronger and stronger. 

Our results can be verified by current experimental setup. By tilting the optical lattice in the axial direction, one can measure Bloch oscillation amplitude and frequency of the itinerant atoms and extract the information of conductivity. Another possible strategy is to detect the dipolar oscillation of the itinerant atoms\cite{Foss2010}, from which the information of conductivity might be extracted as well.  Utilizing the CIR technique, one can also quench the system from ferromagnetic regime to antiferromagnetic regime to study the non-equilibrium dynamics of  Kondo effect\cite{Ashida2018,Ashida2018B}.  Though for simplicity, we take the 1D Kondo model as an example, the approach illustrated in present manuscript can be extended to the high dimensional Kondo model\cite{Yao2018}.

\begin{acknowledgements} 
We are indebted to Hui Zhai, Peng Zhang and Wei Zheng for stimulating discussions. This work is supported by the National Key R\&D  Program of China Grant No. 2018YFA0307601, NSFC Grant No. 11804268.
\end{acknowledgements}

\bibliography{references}

\begin{thebibliography}{34}%
\makeatletter
\providecommand \@ifxundefined [1]{%
 \@ifx{#1\undefined}
}%
\providecommand \@ifnum [1]{%
 \ifnum #1\expandafter \@firstoftwo
 \else \expandafter \@secondoftwo
 \fi
}%
\providecommand \@ifx [1]{%
 \ifx #1\expandafter \@firstoftwo
 \else \expandafter \@secondoftwo
 \fi
}%
\providecommand \natexlab [1]{#1}%
\providecommand \enquote  [1]{``#1''}%
\providecommand \bibnamefont  [1]{#1}%
\providecommand \bibfnamefont [1]{#1}%
\providecommand \citenamefont [1]{#1}%
\providecommand \href@noop [0]{\@secondoftwo}%
\providecommand \href [0]{\begingroup \@sanitize@url \@href}%
\providecommand \@href[1]{\@@startlink{#1}\@@href}%
\providecommand \@@href[1]{\endgroup#1\@@endlink}%
\providecommand \@sanitize@url [0]{\catcode `\\12\catcode `\$12\catcode
  `\&12\catcode `\#12\catcode `\^12\catcode `\_12\catcode `\%12\relax}%
\providecommand \@@startlink[1]{}%
\providecommand \@@endlink[0]{}%
\providecommand \url  [0]{\begingroup\@sanitize@url \@url }%
\providecommand \@url [1]{\endgroup\@href {#1}{\urlprefix }}%
\providecommand \urlprefix  [0]{URL }%
\providecommand \Eprint [0]{\href }%
\providecommand \doibase [0]{http://dx.doi.org/}%
\providecommand \selectlanguage [0]{\@gobble}%
\providecommand \bibinfo  [0]{\@secondoftwo}%
\providecommand \bibfield  [0]{\@secondoftwo}%
\providecommand \translation [1]{[#1]}%
\providecommand \BibitemOpen [0]{}%
\providecommand \bibitemStop [0]{}%
\providecommand \bibitemNoStop [0]{.\EOS\space}%
\providecommand \EOS [0]{\spacefactor3000\relax}%
\providecommand \BibitemShut  [1]{\csname bibitem#1\endcsname}%
\let\auto@bib@innerbib\@empty
\bibitem [{\citenamefont {Bloch}\ \emph {et~al.}(2008)\citenamefont {Bloch},
  \citenamefont {Dalibard},\ and\ \citenamefont {Zwerger}}]{qs1}%
  \BibitemOpen
  \bibfield  {author} {\bibinfo {author} {\bibfnamefont {I.}~\bibnamefont
  {Bloch}}, \bibinfo {author} {\bibfnamefont {J.}~\bibnamefont {Dalibard}}, \
  and\ \bibinfo {author} {\bibfnamefont {W.}~\bibnamefont {Zwerger}},\ }\href
  {\doibase 10.1103/RevModPhys.80.885} {\bibfield  {journal} {\bibinfo
  {journal} {Rev. Mod. Phys.}\ }\textbf {\bibinfo {volume} {80}},\ \bibinfo
  {pages} {885} (\bibinfo {year} {2008})}\BibitemShut {NoStop}%
\bibitem [{\citenamefont {Goldman}\ \emph {et~al.}(2016)\citenamefont
  {Goldman}, \citenamefont {Budich},\ and\ \citenamefont {Zoller}}]{qs2}%
  \BibitemOpen
  \bibfield  {author} {\bibinfo {author} {\bibfnamefont {N.}~\bibnamefont
  {Goldman}}, \bibinfo {author} {\bibfnamefont {J.~C.}\ \bibnamefont {Budich}},
  \ and\ \bibinfo {author} {\bibfnamefont {P.}~\bibnamefont {Zoller}},\ }\href
  {http://dx.doi.org/10.1038/nphys3803} {\bibfield  {journal} {\bibinfo
  {journal} {Nature Physics}\ }\textbf {\bibinfo {volume} {12}},\ \bibinfo
  {pages} {639 EP } (\bibinfo {year} {2016})}\BibitemShut {NoStop}%
\bibitem [{\citenamefont {Gross}\ and\ \citenamefont {Bloch}(2017)}]{qs3}%
  \BibitemOpen
  \bibfield  {author} {\bibinfo {author} {\bibfnamefont {C.}~\bibnamefont
  {Gross}}\ and\ \bibinfo {author} {\bibfnamefont {I.}~\bibnamefont {Bloch}},\
  }\href {\doibase 10.1126/science.aal3837} {\bibfield  {journal} {\bibinfo
  {journal} {Science}\ }\textbf {\bibinfo {volume} {357}},\ \bibinfo {pages}
  {995} (\bibinfo {year} {2017})}\BibitemShut {NoStop}%
\bibitem [{\citenamefont {Regal}\ \emph {et~al.}(2004)\citenamefont {Regal},
  \citenamefont {Greiner},\ and\ \citenamefont {Jin}}]{exp1}%
  \BibitemOpen
  \bibfield  {author} {\bibinfo {author} {\bibfnamefont {C.~A.}\ \bibnamefont
  {Regal}}, \bibinfo {author} {\bibfnamefont {M.}~\bibnamefont {Greiner}}, \
  and\ \bibinfo {author} {\bibfnamefont {D.~S.}\ \bibnamefont {Jin}},\ }\href
  {\doibase 10.1103/PhysRevLett.92.040403} {\bibfield  {journal} {\bibinfo
  {journal} {Phys. Rev. Lett.}\ }\textbf {\bibinfo {volume} {92}},\ \bibinfo
  {pages} {040403} (\bibinfo {year} {2004})}\BibitemShut {NoStop}%
\bibitem [{\citenamefont {Zwierlein}\ \emph {et~al.}(2004)\citenamefont
  {Zwierlein}, \citenamefont {Stan}, \citenamefont {Schunck}, \citenamefont
  {Raupach}, \citenamefont {Kerman},\ and\ \citenamefont {Ketterle}}]{exp2}%
  \BibitemOpen
  \bibfield  {author} {\bibinfo {author} {\bibfnamefont {M.~W.}\ \bibnamefont
  {Zwierlein}}, \bibinfo {author} {\bibfnamefont {C.~A.}\ \bibnamefont {Stan}},
  \bibinfo {author} {\bibfnamefont {C.~H.}\ \bibnamefont {Schunck}}, \bibinfo
  {author} {\bibfnamefont {S.~M.~F.}\ \bibnamefont {Raupach}}, \bibinfo
  {author} {\bibfnamefont {A.~J.}\ \bibnamefont {Kerman}}, \ and\ \bibinfo
  {author} {\bibfnamefont {W.}~\bibnamefont {Ketterle}},\ }\href {\doibase
  10.1103/PhysRevLett.92.120403} {\bibfield  {journal} {\bibinfo  {journal}
  {Phys. Rev. Lett.}\ }\textbf {\bibinfo {volume} {92}},\ \bibinfo {pages}
  {120403} (\bibinfo {year} {2004})}\BibitemShut {NoStop}%
\bibitem [{\citenamefont {Kinast}\ \emph {et~al.}(2004)\citenamefont {Kinast},
  \citenamefont {Hemmer}, \citenamefont {Gehm}, \citenamefont {Turlapov},\ and\
  \citenamefont {Thomas}}]{exp3}%
  \BibitemOpen
  \bibfield  {author} {\bibinfo {author} {\bibfnamefont {J.}~\bibnamefont
  {Kinast}}, \bibinfo {author} {\bibfnamefont {S.~L.}\ \bibnamefont {Hemmer}},
  \bibinfo {author} {\bibfnamefont {M.~E.}\ \bibnamefont {Gehm}}, \bibinfo
  {author} {\bibfnamefont {A.}~\bibnamefont {Turlapov}}, \ and\ \bibinfo
  {author} {\bibfnamefont {J.~E.}\ \bibnamefont {Thomas}},\ }\href {\doibase
  10.1103/PhysRevLett.92.150402} {\bibfield  {journal} {\bibinfo  {journal}
  {Phys. Rev. Lett.}\ }\textbf {\bibinfo {volume} {92}},\ \bibinfo {pages}
  {150402} (\bibinfo {year} {2004})}\BibitemShut {NoStop}%
\bibitem [{\citenamefont {Bourdel}\ \emph {et~al.}(2004)\citenamefont
  {Bourdel}, \citenamefont {Khaykovich}, \citenamefont {Cubizolles},
  \citenamefont {Zhang}, \citenamefont {Chevy}, \citenamefont {Teichmann},
  \citenamefont {Tarruell}, \citenamefont {Kokkelmans},\ and\ \citenamefont
  {Salomon}}]{exp4}%
  \BibitemOpen
  \bibfield  {author} {\bibinfo {author} {\bibfnamefont {T.}~\bibnamefont
  {Bourdel}}, \bibinfo {author} {\bibfnamefont {L.}~\bibnamefont {Khaykovich}},
  \bibinfo {author} {\bibfnamefont {J.}~\bibnamefont {Cubizolles}}, \bibinfo
  {author} {\bibfnamefont {J.}~\bibnamefont {Zhang}}, \bibinfo {author}
  {\bibfnamefont {F.}~\bibnamefont {Chevy}}, \bibinfo {author} {\bibfnamefont
  {M.}~\bibnamefont {Teichmann}}, \bibinfo {author} {\bibfnamefont
  {L.}~\bibnamefont {Tarruell}}, \bibinfo {author} {\bibfnamefont {S.~J. J.
  M.~F.}\ \bibnamefont {Kokkelmans}}, \ and\ \bibinfo {author} {\bibfnamefont
  {C.}~\bibnamefont {Salomon}},\ }\href {\doibase
  10.1103/PhysRevLett.93.050401} {\bibfield  {journal} {\bibinfo  {journal}
  {Phys. Rev. Lett.}\ }\textbf {\bibinfo {volume} {93}},\ \bibinfo {pages}
  {050401} (\bibinfo {year} {2004})}\BibitemShut {NoStop}%
\bibitem [{\citenamefont {Chin}\ \emph {et~al.}(2004)\citenamefont {Chin},
  \citenamefont {Bartenstein}, \citenamefont {Altmeyer}, \citenamefont {Riedl},
  \citenamefont {Jochim}, \citenamefont {Denschlag},\ and\ \citenamefont
  {Grimm}}]{exp5}%
  \BibitemOpen
  \bibfield  {author} {\bibinfo {author} {\bibfnamefont {C.}~\bibnamefont
  {Chin}}, \bibinfo {author} {\bibfnamefont {M.}~\bibnamefont {Bartenstein}},
  \bibinfo {author} {\bibfnamefont {A.}~\bibnamefont {Altmeyer}}, \bibinfo
  {author} {\bibfnamefont {S.}~\bibnamefont {Riedl}}, \bibinfo {author}
  {\bibfnamefont {S.}~\bibnamefont {Jochim}}, \bibinfo {author} {\bibfnamefont
  {J.~H.}\ \bibnamefont {Denschlag}}, \ and\ \bibinfo {author} {\bibfnamefont
  {R.}~\bibnamefont {Grimm}},\ }\href {\doibase 10.1126/science.1100818}
  {\bibfield  {journal} {\bibinfo  {journal} {Science}\ }\textbf {\bibinfo
  {volume} {305}},\ \bibinfo {pages} {1128} (\bibinfo {year}
  {2004})}\BibitemShut {NoStop}%
\bibitem [{\citenamefont {Partridge}\ \emph {et~al.}(2005)\citenamefont
  {Partridge}, \citenamefont {Strecker}, \citenamefont {Kamar}, \citenamefont
  {Jack},\ and\ \citenamefont {Hulet}}]{exp6}%
  \BibitemOpen
  \bibfield  {author} {\bibinfo {author} {\bibfnamefont {G.~B.}\ \bibnamefont
  {Partridge}}, \bibinfo {author} {\bibfnamefont {K.~E.}\ \bibnamefont
  {Strecker}}, \bibinfo {author} {\bibfnamefont {R.~I.}\ \bibnamefont {Kamar}},
  \bibinfo {author} {\bibfnamefont {M.~W.}\ \bibnamefont {Jack}}, \ and\
  \bibinfo {author} {\bibfnamefont {R.~G.}\ \bibnamefont {Hulet}},\ }\href
  {\doibase 10.1103/PhysRevLett.95.020404} {\bibfield  {journal} {\bibinfo
  {journal} {Phys. Rev. Lett.}\ }\textbf {\bibinfo {volume} {95}},\ \bibinfo
  {pages} {020404} (\bibinfo {year} {2005})}\BibitemShut {NoStop}%
\bibitem [{\citenamefont {Zwierlein}\ \emph {et~al.}(2005)\citenamefont
  {Zwierlein}, \citenamefont {Abo-Shaeer}, \citenamefont {Schirotzek},
  \citenamefont {Schunck},\ and\ \citenamefont {Ketterle}}]{exp7}%
  \BibitemOpen
  \bibfield  {author} {\bibinfo {author} {\bibfnamefont {M.~W.}\ \bibnamefont
  {Zwierlein}}, \bibinfo {author} {\bibfnamefont {J.~R.}\ \bibnamefont
  {Abo-Shaeer}}, \bibinfo {author} {\bibfnamefont {A.}~\bibnamefont
  {Schirotzek}}, \bibinfo {author} {\bibfnamefont {C.~H.}\ \bibnamefont
  {Schunck}}, \ and\ \bibinfo {author} {\bibfnamefont {W.}~\bibnamefont
  {Ketterle}},\ }\href {http://dx.doi.org/10.1038/nature03858} {\bibfield
  {journal} {\bibinfo  {journal} {Nature}\ }\textbf {\bibinfo {volume} {435}},\
  \bibinfo {pages} {1047 EP } (\bibinfo {year} {2005})}\BibitemShut {NoStop}%
\bibitem [{\citenamefont {Falco}\ \emph {et~al.}(2004)\citenamefont {Falco},
  \citenamefont {Duine},\ and\ \citenamefont {Stoof}}]{stoof2004}%
  \BibitemOpen
  \bibfield  {author} {\bibinfo {author} {\bibfnamefont {G.~M.}\ \bibnamefont
  {Falco}}, \bibinfo {author} {\bibfnamefont {R.~A.}\ \bibnamefont {Duine}}, \
  and\ \bibinfo {author} {\bibfnamefont {H.~T.~C.}\ \bibnamefont {Stoof}},\
  }\href {\doibase 10.1103/PhysRevLett.92.140402} {\bibfield  {journal}
  {\bibinfo  {journal} {Phys. Rev. Lett.}\ }\textbf {\bibinfo {volume} {92}},\
  \bibinfo {pages} {140402} (\bibinfo {year} {2004})}\BibitemShut {NoStop}%
\bibitem [{\citenamefont {Gorshkov}\ \emph {et~al.}(2010)\citenamefont
  {Gorshkov}, \citenamefont {Hermele}, \citenamefont {Gurarie}, \citenamefont
  {Xu}, \citenamefont {Julienne}, \citenamefont {Ye}, \citenamefont {Zoller},
  \citenamefont {Demler}, \citenamefont {Lukin},\ and\ \citenamefont
  {Rey}}]{Gorshkov2010}%
  \BibitemOpen
  \bibfield  {author} {\bibinfo {author} {\bibfnamefont {A.~V.}\ \bibnamefont
  {Gorshkov}}, \bibinfo {author} {\bibfnamefont {M.}~\bibnamefont {Hermele}},
  \bibinfo {author} {\bibfnamefont {V.}~\bibnamefont {Gurarie}}, \bibinfo
  {author} {\bibfnamefont {C.}~\bibnamefont {Xu}}, \bibinfo {author}
  {\bibfnamefont {P.~S.}\ \bibnamefont {Julienne}}, \bibinfo {author}
  {\bibfnamefont {J.}~\bibnamefont {Ye}}, \bibinfo {author} {\bibfnamefont
  {P.}~\bibnamefont {Zoller}}, \bibinfo {author} {\bibfnamefont
  {E.}~\bibnamefont {Demler}}, \bibinfo {author} {\bibfnamefont {M.~D.}\
  \bibnamefont {Lukin}}, \ and\ \bibinfo {author} {\bibfnamefont {A.~M.}\
  \bibnamefont {Rey}},\ }\href {http://dx.doi.org/10.1038/nphys1535
  http://10.0.4.14/nphys1535
  https://www.nature.com/articles/nphys1535{\#}supplementary-information}
  {\bibfield  {journal} {\bibinfo  {journal} {Nature Physics}\ }\textbf
  {\bibinfo {volume} {6}},\ \bibinfo {pages} {289} (\bibinfo {year}
  {2010})}\BibitemShut {NoStop}%
\bibitem [{\citenamefont {Foss-Feig}\ \emph {et~al.}(2010)\citenamefont
  {Foss-Feig}, \citenamefont {Hermele},\ and\ \citenamefont {Rey}}]{Foss2010}%
  \BibitemOpen
  \bibfield  {author} {\bibinfo {author} {\bibfnamefont {M.}~\bibnamefont
  {Foss-Feig}}, \bibinfo {author} {\bibfnamefont {M.}~\bibnamefont {Hermele}},
  \ and\ \bibinfo {author} {\bibfnamefont {A.~M.}\ \bibnamefont {Rey}},\ }\href
  {\doibase 10.1103/PhysRevA.81.051603} {\bibfield  {journal} {\bibinfo
  {journal} {Phys. Rev. A}\ }\textbf {\bibinfo {volume} {81}},\ \bibinfo
  {pages} {051603} (\bibinfo {year} {2010})}\BibitemShut {NoStop}%
\bibitem [{\citenamefont {Bauer}\ \emph {et~al.}(2013)\citenamefont {Bauer},
  \citenamefont {Salomon},\ and\ \citenamefont {Demler}}]{Bauer2013}%
  \BibitemOpen
  \bibfield  {author} {\bibinfo {author} {\bibfnamefont {J.}~\bibnamefont
  {Bauer}}, \bibinfo {author} {\bibfnamefont {C.}~\bibnamefont {Salomon}}, \
  and\ \bibinfo {author} {\bibfnamefont {E.}~\bibnamefont {Demler}},\ }\href
  {\doibase 10.1103/PhysRevLett.111.215304} {\bibfield  {journal} {\bibinfo
  {journal} {Phys. Rev. Lett.}\ }\textbf {\bibinfo {volume} {111}},\ \bibinfo
  {pages} {215304} (\bibinfo {year} {2013})}\BibitemShut {NoStop}%
\bibitem [{\citenamefont {Nishida}(2013)}]{nishida2013}%
  \BibitemOpen
  \bibfield  {author} {\bibinfo {author} {\bibfnamefont {Y.}~\bibnamefont
  {Nishida}},\ }\href {\doibase 10.1103/PhysRevLett.111.135301} {\bibfield
  {journal} {\bibinfo  {journal} {Phys. Rev. Lett.}\ }\textbf {\bibinfo
  {volume} {111}},\ \bibinfo {pages} {135301} (\bibinfo {year}
  {2013})}\BibitemShut {NoStop}%
\bibitem [{\citenamefont {Isaev}\ and\ \citenamefont {Rey}(2015)}]{Isaev2015}%
  \BibitemOpen
  \bibfield  {author} {\bibinfo {author} {\bibfnamefont {L.}~\bibnamefont
  {Isaev}}\ and\ \bibinfo {author} {\bibfnamefont {A.~M.}\ \bibnamefont
  {Rey}},\ }\href {\doibase 10.1103/PhysRevLett.115.165302} {\bibfield
  {journal} {\bibinfo  {journal} {Phys. Rev. Lett.}\ }\textbf {\bibinfo
  {volume} {115}},\ \bibinfo {pages} {165302} (\bibinfo {year}
  {2015})}\BibitemShut {NoStop}%
\bibitem [{\citenamefont {Kuzmenko}\ \emph {et~al.}(2015)\citenamefont
  {Kuzmenko}, \citenamefont {Kuzmenko}, \citenamefont {Avishai},\ and\
  \citenamefont {Kikoin}}]{kuzmenko2015}%
  \BibitemOpen
  \bibfield  {author} {\bibinfo {author} {\bibfnamefont {I.}~\bibnamefont
  {Kuzmenko}}, \bibinfo {author} {\bibfnamefont {T.}~\bibnamefont {Kuzmenko}},
  \bibinfo {author} {\bibfnamefont {Y.}~\bibnamefont {Avishai}}, \ and\
  \bibinfo {author} {\bibfnamefont {K.}~\bibnamefont {Kikoin}},\ }\href
  {\doibase 10.1103/PhysRevB.91.165131} {\bibfield  {journal} {\bibinfo
  {journal} {Phys. Rev. B}\ }\textbf {\bibinfo {volume} {91}},\ \bibinfo
  {pages} {165131} (\bibinfo {year} {2015})}\BibitemShut {NoStop}%
\bibitem [{\citenamefont {Nishida}(2016)}]{nishida2016}%
  \BibitemOpen
  \bibfield  {author} {\bibinfo {author} {\bibfnamefont {Y.}~\bibnamefont
  {Nishida}},\ }\href {\doibase 10.1103/PhysRevA.93.011606} {\bibfield
  {journal} {\bibinfo  {journal} {Phys. Rev. A}\ }\textbf {\bibinfo {volume}
  {93}},\ \bibinfo {pages} {011606} (\bibinfo {year} {2016})}\BibitemShut
  {NoStop}%
\bibitem [{\citenamefont {Zhang}\ \emph {et~al.}(2016)\citenamefont {Zhang},
  \citenamefont {Zhang}, \citenamefont {Cheng}, \citenamefont {Chen},
  \citenamefont {Zhang},\ and\ \citenamefont {Zhai}}]{ZR2016}%
  \BibitemOpen
  \bibfield  {author} {\bibinfo {author} {\bibfnamefont {R.}~\bibnamefont
  {Zhang}}, \bibinfo {author} {\bibfnamefont {D.}~\bibnamefont {Zhang}},
  \bibinfo {author} {\bibfnamefont {Y.}~\bibnamefont {Cheng}}, \bibinfo
  {author} {\bibfnamefont {W.}~\bibnamefont {Chen}}, \bibinfo {author}
  {\bibfnamefont {P.}~\bibnamefont {Zhang}}, \ and\ \bibinfo {author}
  {\bibfnamefont {H.}~\bibnamefont {Zhai}},\ }\href {\doibase
  10.1103/PhysRevA.93.043601} {\bibfield  {journal} {\bibinfo  {journal} {Phys.
  Rev. A}\ }\textbf {\bibinfo {volume} {93}},\ \bibinfo {pages} {043601}
  (\bibinfo {year} {2016})}\BibitemShut {NoStop}%
\bibitem [{\citenamefont {Kuzmenko}\ \emph {et~al.}(2018)\citenamefont
  {Kuzmenko}, \citenamefont {Kuzmenko}, \citenamefont {Avishai},\ and\
  \citenamefont {Jo}}]{Kuzemenko2018}%
  \BibitemOpen
  \bibfield  {author} {\bibinfo {author} {\bibfnamefont {I.}~\bibnamefont
  {Kuzmenko}}, \bibinfo {author} {\bibfnamefont {T.}~\bibnamefont {Kuzmenko}},
  \bibinfo {author} {\bibfnamefont {Y.}~\bibnamefont {Avishai}}, \ and\
  \bibinfo {author} {\bibfnamefont {G.-B.}\ \bibnamefont {Jo}},\ }\href
  {\doibase 10.1103/PhysRevB.97.075124} {\bibfield  {journal} {\bibinfo
  {journal} {Phys. Rev. B}\ }\textbf {\bibinfo {volume} {97}},\ \bibinfo
  {pages} {075124} (\bibinfo {year} {2018})}\BibitemShut {NoStop}%
\bibitem [{\citenamefont {Cheng}\ \emph {et~al.}(2017)\citenamefont {Cheng},
  \citenamefont {Zhang}, \citenamefont {Zhang},\ and\ \citenamefont
  {Zhai}}]{1dcir}%
  \BibitemOpen
  \bibfield  {author} {\bibinfo {author} {\bibfnamefont {Y.}~\bibnamefont
  {Cheng}}, \bibinfo {author} {\bibfnamefont {R.}~\bibnamefont {Zhang}},
  \bibinfo {author} {\bibfnamefont {P.}~\bibnamefont {Zhang}}, \ and\ \bibinfo
  {author} {\bibfnamefont {H.}~\bibnamefont {Zhai}},\ }\href {\doibase
  10.1103/PhysRevA.96.063605} {\bibfield  {journal} {\bibinfo  {journal} {Phys.
  Rev. A}\ }\textbf {\bibinfo {volume} {96}},\ \bibinfo {pages} {063605}
  (\bibinfo {year} {2017})}\BibitemShut {NoStop}%
\bibitem [{\citenamefont {Yao}\ \emph {et~al.}()\citenamefont {Yao},
  \citenamefont {Zhai},\ and\ \citenamefont {Zhang}}]{Yao2018}%
  \BibitemOpen
  \bibfield  {author} {\bibinfo {author} {\bibfnamefont {J.}~\bibnamefont
  {Yao}}, \bibinfo {author} {\bibfnamefont {H.}~\bibnamefont {Zhai}}, \ and\
  \bibinfo {author} {\bibfnamefont {R.}~\bibnamefont {Zhang}},\ }\href@noop {}
  {\ }\Eprint {http://arxiv.org/abs/1809.06340} {arXiv:1809.06340} \BibitemShut
  {NoStop}%
\bibitem [{\citenamefont {Anderson}(1970)}]{poor}%
  \BibitemOpen
  \bibfield  {author} {\bibinfo {author} {\bibfnamefont {P.~W.}\ \bibnamefont
  {Anderson}},\ }\href {http://stacks.iop.org/0022-3719/3/i=12/a=008}
  {\bibfield  {journal} {\bibinfo  {journal} {Journal of Physics C: Solid State
  Physics}\ }\textbf {\bibinfo {volume} {3}},\ \bibinfo {pages} {2436}
  (\bibinfo {year} {1970})}\BibitemShut {NoStop}%
\bibitem [{\citenamefont {Kondo}(1964)}]{ori}%
  \BibitemOpen
  \bibfield  {author} {\bibinfo {author} {\bibfnamefont {J.}~\bibnamefont
  {Kondo}},\ }\href {\doibase 10.1143/PTP.32.37} {\bibfield  {journal}
  {\bibinfo  {journal} {Progress of Theoretical Physics}\ }\textbf {\bibinfo
  {volume} {32}},\ \bibinfo {pages} {37} (\bibinfo {year} {1964})}\BibitemShut
  {NoStop}%
\bibitem [{\citenamefont {Barber}\ \emph {et~al.}(2008)\citenamefont {Barber},
  \citenamefont {Stalnaker}, \citenamefont {Lemke}, \citenamefont {Poli},
  \citenamefont {Oates}, \citenamefont {Fortier}, \citenamefont {Diddams},
  \citenamefont {Hollberg}, \citenamefont {Hoyt}, \citenamefont
  {Taichenachev},\ and\ \citenamefont {Yudin}}]{mag1}%
  \BibitemOpen
  \bibfield  {author} {\bibinfo {author} {\bibfnamefont {Z.~W.}\ \bibnamefont
  {Barber}}, \bibinfo {author} {\bibfnamefont {J.~E.}\ \bibnamefont
  {Stalnaker}}, \bibinfo {author} {\bibfnamefont {N.~D.}\ \bibnamefont
  {Lemke}}, \bibinfo {author} {\bibfnamefont {N.}~\bibnamefont {Poli}},
  \bibinfo {author} {\bibfnamefont {C.~W.}\ \bibnamefont {Oates}}, \bibinfo
  {author} {\bibfnamefont {T.~M.}\ \bibnamefont {Fortier}}, \bibinfo {author}
  {\bibfnamefont {S.~A.}\ \bibnamefont {Diddams}}, \bibinfo {author}
  {\bibfnamefont {L.}~\bibnamefont {Hollberg}}, \bibinfo {author}
  {\bibfnamefont {C.~W.}\ \bibnamefont {Hoyt}}, \bibinfo {author}
  {\bibfnamefont {A.~V.}\ \bibnamefont {Taichenachev}}, \ and\ \bibinfo
  {author} {\bibfnamefont {V.~I.}\ \bibnamefont {Yudin}},\ }\href {\doibase
  10.1103/PhysRevLett.100.103002} {\bibfield  {journal} {\bibinfo  {journal}
  {Phys. Rev. Lett.}\ }\textbf {\bibinfo {volume} {100}},\ \bibinfo {pages}
  {103002} (\bibinfo {year} {2008})}\BibitemShut {NoStop}%
\bibitem [{\citenamefont {Dzuba}\ and\ \citenamefont
  {Derevianko}(2010)}]{mag2}%
  \BibitemOpen
  \bibfield  {author} {\bibinfo {author} {\bibfnamefont {V.~A.}\ \bibnamefont
  {Dzuba}}\ and\ \bibinfo {author} {\bibfnamefont {A.}~\bibnamefont
  {Derevianko}},\ }\href {http://stacks.iop.org/0953-4075/43/i=7/a=074011}
  {\bibfield  {journal} {\bibinfo  {journal} {Journal of Physics B: Atomic,
  Molecular and Optical Physics}\ }\textbf {\bibinfo {volume} {43}},\ \bibinfo
  {pages} {074011} (\bibinfo {year} {2010})}\BibitemShut {NoStop}%
\bibitem [{\citenamefont {Cappellini}\ \emph {et~al.}(2014)\citenamefont
  {Cappellini}, \citenamefont {Mancini}, \citenamefont {Pagano}, \citenamefont
  {Lombardi}, \citenamefont {Livi}, \citenamefont {Siciliani~de Cumis},
  \citenamefont {Cancio}, \citenamefont {Pizzocaro}, \citenamefont {Calonico},
  \citenamefont {Levi}, \citenamefont {Sias}, \citenamefont {Catani},
  \citenamefont {Inguscio},\ and\ \citenamefont {Fallani}}]{spinexchange1}%
  \BibitemOpen
  \bibfield  {author} {\bibinfo {author} {\bibfnamefont {G.}~\bibnamefont
  {Cappellini}}, \bibinfo {author} {\bibfnamefont {M.}~\bibnamefont {Mancini}},
  \bibinfo {author} {\bibfnamefont {G.}~\bibnamefont {Pagano}}, \bibinfo
  {author} {\bibfnamefont {P.}~\bibnamefont {Lombardi}}, \bibinfo {author}
  {\bibfnamefont {L.}~\bibnamefont {Livi}}, \bibinfo {author} {\bibfnamefont
  {M.}~\bibnamefont {Siciliani~de Cumis}}, \bibinfo {author} {\bibfnamefont
  {P.}~\bibnamefont {Cancio}}, \bibinfo {author} {\bibfnamefont
  {M.}~\bibnamefont {Pizzocaro}}, \bibinfo {author} {\bibfnamefont
  {D.}~\bibnamefont {Calonico}}, \bibinfo {author} {\bibfnamefont
  {F.}~\bibnamefont {Levi}}, \bibinfo {author} {\bibfnamefont {C.}~\bibnamefont
  {Sias}}, \bibinfo {author} {\bibfnamefont {J.}~\bibnamefont {Catani}},
  \bibinfo {author} {\bibfnamefont {M.}~\bibnamefont {Inguscio}}, \ and\
  \bibinfo {author} {\bibfnamefont {L.}~\bibnamefont {Fallani}},\ }\href
  {\doibase 10.1103/PhysRevLett.113.120402} {\bibfield  {journal} {\bibinfo
  {journal} {Phys. Rev. Lett.}\ }\textbf {\bibinfo {volume} {113}},\ \bibinfo
  {pages} {120402} (\bibinfo {year} {2014})}\BibitemShut {NoStop}%
\bibitem [{\citenamefont {Scazza}\ \emph {et~al.}(2014)\citenamefont {Scazza},
  \citenamefont {Hofrichter}, \citenamefont {H{\"{o}}fer}, \citenamefont {{De
  Groot}}, \citenamefont {Bloch},\ and\ \citenamefont
  {F{\"{o}}lling}}]{spinexchange2}%
  \BibitemOpen
  \bibfield  {author} {\bibinfo {author} {\bibfnamefont {F.}~\bibnamefont
  {Scazza}}, \bibinfo {author} {\bibfnamefont {C.}~\bibnamefont {Hofrichter}},
  \bibinfo {author} {\bibfnamefont {M.}~\bibnamefont {H{\"{o}}fer}}, \bibinfo
  {author} {\bibfnamefont {P.~C.}\ \bibnamefont {{De Groot}}}, \bibinfo
  {author} {\bibfnamefont {I.}~\bibnamefont {Bloch}}, \ and\ \bibinfo {author}
  {\bibfnamefont {S.}~\bibnamefont {F{\"{o}}lling}},\ }\href {\doibase
  10.1038/nphys3061} {\bibfield  {journal} {\bibinfo  {journal} {Nature
  Physics}\ } (\bibinfo {year} {2014}),\ 10.1038/nphys3061},\ \Eprint
  {http://arxiv.org/abs/1403.4761} {arXiv:1403.4761} \BibitemShut {NoStop}%
\bibitem [{\citenamefont {Riegger}\ \emph {et~al.}(2018)\citenamefont
  {Riegger}, \citenamefont {Darkwah~Oppong}, \citenamefont {H\"ofer},
  \citenamefont {Fernandes}, \citenamefont {Bloch},\ and\ \citenamefont
  {F\"olling}}]{1dcirex}%
  \BibitemOpen
  \bibfield  {author} {\bibinfo {author} {\bibfnamefont {L.}~\bibnamefont
  {Riegger}}, \bibinfo {author} {\bibfnamefont {N.}~\bibnamefont
  {Darkwah~Oppong}}, \bibinfo {author} {\bibfnamefont {M.}~\bibnamefont
  {H\"ofer}}, \bibinfo {author} {\bibfnamefont {D.~R.}\ \bibnamefont
  {Fernandes}}, \bibinfo {author} {\bibfnamefont {I.}~\bibnamefont {Bloch}}, \
  and\ \bibinfo {author} {\bibfnamefont {S.}~\bibnamefont {F\"olling}},\ }\href
  {\doibase 10.1103/PhysRevLett.120.143601} {\bibfield  {journal} {\bibinfo
  {journal} {Phys. Rev. Lett.}\ }\textbf {\bibinfo {volume} {120}},\ \bibinfo
  {pages} {143601} (\bibinfo {year} {2018})}\BibitemShut {NoStop}%
\bibitem [{\citenamefont {Zhang}\ and\ \citenamefont {Zhang}()}]{Zhang2018}%
  \BibitemOpen
  \bibfield  {author} {\bibinfo {author} {\bibfnamefont {R.}~\bibnamefont
  {Zhang}}\ and\ \bibinfo {author} {\bibfnamefont {P.}~\bibnamefont {Zhang}},\
  }\href {http://arxiv.org/abs/1804.03388} {\ }\Eprint
  {http://arxiv.org/abs/1804.03388} {arXiv:1804.03388} \BibitemShut {NoStop}%
\bibitem [{\citenamefont {Hewson}(1993)}]{kondo}%
  \BibitemOpen
  \bibfield  {author} {\bibinfo {author} {\bibfnamefont {A.~C.}\ \bibnamefont
  {Hewson}},\ }\href {\doibase 10.1017/CBO9780511470752} {\emph {\bibinfo
  {title} {The Kondo Problem to Heavy Fermions}}},\ Cambridge Studies in
  Magnetism\ (\bibinfo  {publisher} {Cambridge University Press},\ \bibinfo
  {year} {1993})\BibitemShut {NoStop}%
\bibitem [{\citenamefont {{Sadovskii}}(2006)}]{dia}%
  \BibitemOpen
  \bibfield  {author} {\bibinfo {author} {\bibfnamefont {M.~V.}\ \bibnamefont
  {{Sadovskii}}},\ }\href {\doibase 10.1142/6011} {\emph {\bibinfo {title}
  {Diagrammatics: Lectures on Selected Problems in Condensed Matter
  Theory.~Edited by SADOVSKII MICHAEL V.~Published by World Scientific
  Publishing Co.~Pte.~Ltd., .~ISBN \#9789812774361}}}\ (\bibinfo  {publisher}
  {World Scientific Publishing Co},\ \bibinfo {year} {2006})\BibitemShut
  {NoStop}%
\bibitem [{\citenamefont {Ashida}\ \emph
  {et~al.}(2018{\natexlab{a}})\citenamefont {Ashida}, \citenamefont {Shi},
  \citenamefont {Ba\~nuls}, \citenamefont {Cirac},\ and\ \citenamefont
  {Demler}}]{Ashida2018}%
  \BibitemOpen
  \bibfield  {author} {\bibinfo {author} {\bibfnamefont {Y.}~\bibnamefont
  {Ashida}}, \bibinfo {author} {\bibfnamefont {T.}~\bibnamefont {Shi}},
  \bibinfo {author} {\bibfnamefont {M.~C.}\ \bibnamefont {Ba\~nuls}}, \bibinfo
  {author} {\bibfnamefont {J.~I.}\ \bibnamefont {Cirac}}, \ and\ \bibinfo
  {author} {\bibfnamefont {E.}~\bibnamefont {Demler}},\ }\href {\doibase
  10.1103/PhysRevLett.121.026805} {\bibfield  {journal} {\bibinfo  {journal}
  {Phys. Rev. Lett.}\ }\textbf {\bibinfo {volume} {121}},\ \bibinfo {pages}
  {026805} (\bibinfo {year} {2018}{\natexlab{a}})}\BibitemShut {NoStop}%
\bibitem [{\citenamefont {Ashida}\ \emph
  {et~al.}(2018{\natexlab{b}})\citenamefont {Ashida}, \citenamefont {Shi},
  \citenamefont {Ba\~nuls}, \citenamefont {Cirac},\ and\ \citenamefont
  {Demler}}]{Ashida2018B}%
  \BibitemOpen
  \bibfield  {author} {\bibinfo {author} {\bibfnamefont {Y.}~\bibnamefont
  {Ashida}}, \bibinfo {author} {\bibfnamefont {T.}~\bibnamefont {Shi}},
  \bibinfo {author} {\bibfnamefont {M.~C.}\ \bibnamefont {Ba\~nuls}}, \bibinfo
  {author} {\bibfnamefont {J.~I.}\ \bibnamefont {Cirac}}, \ and\ \bibinfo
  {author} {\bibfnamefont {E.}~\bibnamefont {Demler}},\ }\href {\doibase
  10.1103/PhysRevB.98.024103} {\bibfield  {journal} {\bibinfo  {journal} {Phys.
  Rev. B}\ }\textbf {\bibinfo {volume} {98}},\ \bibinfo {pages} {024103}
  (\bibinfo {year} {2018}{\natexlab{b}})}\BibitemShut {NoStop}%
\end{thebibliography}%

\end{document}